\documentclass[twocolumn]{aastex631}
\usepackage{multirow}
\shorttitle{Molecular K–S Relation Across CO Transitions}

\shortauthors{Samboco \& Keenan}
\graphicspath{{./}{figures/}}

\begin{document}

 \title{Constraining the Molecular Kennicutt–Schmidt Relation with Multi-Transition CO Observations of Nearby Galaxies}

\author[0000-0001-7240-5152]{Victória da G. G. Samboco}
\email{vicky.samboco@gmail.com}
\affiliation{Centre for Radio Astronomy Techniques and Technologies (RATT), Department of Physics and Electronics, Rhodes University, Makhanda, 6140, South Africa \\}
\affiliation{Associação Moçambicana de Astronomia (AMAS), Maputo, Mozambique}
\affiliation{Max-Planck Institute for Astronomy, Königstuhl 17, 69117 Heidelberg, Germany}

\author[0000-0003-1859-9640]{Ryan Keenan}
\email{keenan@mpia.de}
\affiliation{Max-Planck Institute for Astronomy, Königstuhl 17, 69117 Heidelberg, Germany}








\begin{abstract}
The relationship between the star formation rate surface density and the molecular gas surface density in galaxies is key to understanding galaxy evolution. To investigate the molecular Kennicutt-Schmidt (K--S) relation and its dependence on gas density, we analyze a uniform sample of 36 nearby galaxies from the AMISS survey, focusing on the CO(1--0), CO(2--1), and CO(3--2) transitions, which trace progressively denser and warmer molecular gas. Using statistical methods that combine binning with Markov Chain Monte Carlo (MCMC) fitting, we derive the slope, scatter, and intercept of the $\Sigma_{\mathrm{SFR}}$--$\Sigma_{\mathrm{CO}}$ relation for each transition. We find power-law slopes of 1.26, 1.14, and 1.07 for CO(1--0), CO(2--1), and CO(3--2), respectively, consistent with a trend toward increasingly linear star formation relations at higher-J transitions.
This behavior supports the idea that denser gas is more directly linked to ongoing star formation and is consistent with previous findings of near-linear correlations between HCN or high-J CO luminosities and global SFR. The observed trend suggests an underlying relation between gas and SFR volume densities with a power-law index of $\sim$1.5, indicating enhanced star formation efficiency in denser environments. These findings underscore the critical role of dense gas in regulating star formation and highlight the importance of tracer selection and excitation conditions when interpreting the K-S relation across different environments.

\end{abstract}

\keywords{\textit{Astronomy Thesaurus concepts}: Interstellar medium (847); Surveys (1671); Molecular gas (1073); Millimeter astronomy (1061)}


\section{Introduction} \label{sec:intro}

Star formation is a fundamental process by which a galaxy converts gas into stars, making it a crucial driver of galaxy evolution. At the center of this process lies the cold Interstellar Medium (ISM), which fuels star formation and drives the global life cycle of a galaxy \citep{saintonge+22}. Within the ISM, molecular hydrogen (H$_2$) is the primary component of the molecular gas that provides the fundamental material for star formation  \citep{mckee2007theory, bigiel2014interstellar}. Understanding its distribution and behavior is key to uncovering the mechanisms that govern galaxy evolution over cosmic time.

Despite hydrogen being the most abundant component of the ISM, molecular H$_2$ is difficult to observe directly due to its symmetric structure, which limits its interaction with light \citep{hollenbach1971molecular, saintonge+22}. To overcome this challenge, astronomers rely on indirect tracers, particularly carbon monoxide ($^{12}$C$^{16}$O, hereafter CO). CO is the second most abundant molecule in the ISM and is widely used to trace molecular gas across various galactic environments \citep{sandstrom2013co, bolatto+13}.

A critical component of understanding galaxy evolution is probing the relationship between the amount and distribution of cold gas and the rate at which star formation occurs. One of the key empirical frameworks developed to explore this connection is the Kennicutt--Schmidt (K--S) relation, which expresses a power-law relation between the surface densities of star formation ($\Sigma_{\rm SFR}$) and neutral gas ($\Sigma_{\rm gas}$). First proposed by \citet{schmidt1959rate} and later expanded by \citet{kennicutt1998global, kennicutt1998star}, the K--S relation provides a foundation for quantifying star formation efficiency and examining how it varies across different galactic environments.

The K--S relation was initially formulated in terms of total gas surface density (H\,\textsc{i} + H$_2$). However, studies such as \citet{Robertson_2008}, \citet{Bigiel_2008}, and \citet{schruba2011molecular} have shown that molecular gas, rather than total gas, is the dominant factor driving star formation. These findings are supported by more recent work, which confirms a strong intrinsic correlation between star formation surface density ($\Sigma_{\rm SFR}$) and molecular gas surface density ($\Sigma_{\rm mol}$), with little to no dependence on stellar mass or atomic gas once molecular gas is accounted for \citep{baker2022almaquest}. In this paper, we refer to the scaling relation between $\Sigma_{\rm SFR}$ and $\Sigma_{\rm mol}$ as the \textit{molecular K--S relation}.
A related concept is the dense gas--star formation relation (also known as the Gao--Solomon relation), a tight and nearly linear (power-law slope $\sim 1$) correlation between SFR and the mass of molecular gas above a certain density threshold. 

Such linear correlations are widely observed using emission lines that can only be excited in dense gas, such as lines from the HCN molecule or CO transitions above CO(4--3) (e.g., \citealt{Gao_2004}, \citealt{Liu_2015}, \citealt{jimenez-donaire+19}, \citealt{neumann+23}, \citealt{lin+24}).

Although the molecular K--S relation has been widely studied, most analyses have focused on a single CO transition, typically CO(1--0), CO(2--1), or CO(3--2), and often use heterogeneous galaxy samples. A handful of studies consider multiple CO transitions together \citep[e.g.,][]{morokuma-matsui2017, lamperti+20, yajima+21}; however, none have systematically examined the global molecular K--S relation using all three of the lowest-energy CO lines observed in the same galaxy sample at once. This is crucial because different CO transitions reveal distinct physical states of the gas, with some transitions highlighting denser, cooler regions, and others detecting warmer, more diffuse molecular gas \citep{sakamoto1999mass, shetty2011modelling, roman2016distribution}. Analyzing multiple lines provides a more comprehensive understanding of how different tracers of the molecular gas conditions influence star formation.

This study investigates the molecular K--S relation using three CO transitions: CO(1--0), CO(2--1), and CO(3--2) in a uniform sample of 36 nearby galaxies from the Arizona Molecular ISM Survey with the Sub-Millimeter Telescope (SMT)  \citep[AMISS;][]{keenan2024arizona}, one of world’s most precise millimeter-wave telescopes, with a surface accuracy of 15 microns (rms), specifically built for observations in the sub-millimeter wavelength spectrum \citep{martin1990submillimeter}. Analyzing multiple lines in a uniform sample provides an ideal platform for investigating the relationship between molecular gas and star formation under consistent observational conditions. We aim to (i) compare the molecular K--S relation derived from different CO transitions, (ii) quantify differences in the slope and scatter of the relation across these transitions, and (iii) interpret these differences in terms of the physical conditions of the molecular gas and their role in the star formation process. A statistical approach, specifically using the Markov Chain Monte Carlo (MCMC) fitting, is employed to determine the K--S slopes and their associated uncertainties for each CO transition.

This paper is organized as follows: Section~\ref{sec:data} describes the dataset and methodology, Section~\ref{sec:results} presents our results, Section~\ref{sec:discussion} discusses their implications, and Section~\ref{sec:conclusion} summarizes our conclusions. Throughout this paper, $\log$ denotes base 10 logarithms. A flat $\Lambda$CDM cosmology with $H_0 = 70$~km~s$^{-1}$~Mpc$^{-1}$ and $\Omega_m = 0.3$ is assumed for computing luminosities. SFRs are derived assuming a \citet{chabrier2003IMF} initial mass function.

\section{Observations and Data Analysis} \label{sec:data}

\begin{figure}
\centering
\includegraphics[width=0.49\textwidth]{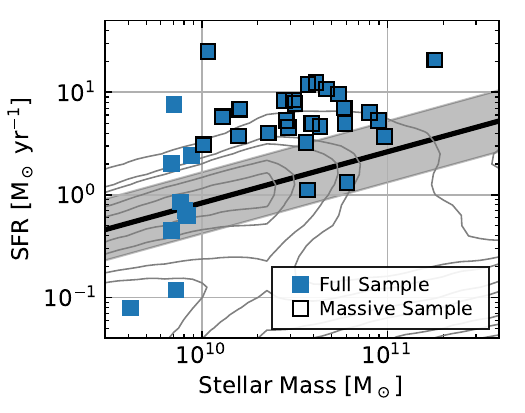}
\caption{The distribution of our galaxy sample in stellar mass and SFR. Our full sample of 36 galaxies is represented by blue boxes. Black outlines denote the subsample of massive, star-forming galaxies (Stellar Mass $\geq10^{10}$ M$_{\odot}$, SFR$\geq1$ M$_{\odot}$yr$^{-1}$) used at points in this work. These are sources detected in CO(1--0), CO(2--1), and CO(3--2). The gray region represents the star forming galaxy main sequence \citep{speagle2014highly}. \label{fig1}}
\end{figure}

The galaxy sample used here (Figure \ref{fig1}) originates from the Arizona Molecular ISM Survey with the SMT \citep[AMISS;][]{keenan2024arizona}. AMISS is a comprehensive survey of multi-CO line emissions from a sample of 176 nearby galaxies at redshifts ranging from 0.02 to 0.05, with a wide range of stellar masses, from $10^9 M_{\odot}$ to $10^{11.5} M_{\odot}$, and SFRs, ranging from $10^{-3}$ to $10^{1.5} M_{\odot}/\text{yr}$. Of the total sample, 45 galaxies were observed in CO(3--2), CO(2--1), and archival CO(1--0), while the remaining 131 galaxies were observed in CO(2--1) and CO(1--0) only.

A detailed description of AMISS can be found in \citet{keenan2024arizona}, and we summarize relevant details here. Observations of CO(2--1) and CO(3--2) were conducted using the Submillimeter Telescope (SMT) and supplemented by archival CO(3--2) observations with the James Clerk Maxwell Telescope (JCMT) from \citep{lamperti+20}. The CO(1--0) data were obtained from the xCOLD GASS survey \citep{saintonge+17} conducted with the IRAM 30m Telescope and new observations with the Kitt Peak 12m Telescope. To reduce systematic uncertainnesses arising from  calibration differences  (\citet{10.1093/mnras/stab859} and \citet{Leroy_2022}), we utilized one telescope per line whenever possible, the IRAM 30 m for CO(1--0) and the Arizona Radio Observatory (AROS)'s Submillimeter Telescope (SMT) for CO(2--1) and CO(3--2) with all observations done for a given line conducted under standardized set of procedures and reduced in a consistent manner.

This paper uses a sample of 36 galaxies with detections of all three CO lines. For galaxies with measurements from both SMT and JCMT, the AMISS survey adopts the value with lower statistical noise. \citet{keenan2024arizona} verified that the aperture-corrected CO(3--2) luminosities are consistent between the two datasets, minimizing potential systematics. At various points we also use a subset of 26 galaxies that excludes low-mass (stellar mass $<10^{10}$ M${\odot}$) and low-SFR (SFR $<1$ M${\odot}$yr$^{-1}$) systems; these galaxies are highlighted by black boxes in Figure~\ref{fig1}.

The beam sizes for the CO observations are 21\arcsec at 115.3 GHz for CO(1--0) (IRAM~30m), 31\arcsec at 230.5 GHz for CO(2--1), and 22\arcsec (SMT) or 14\arcsec (JCMT) at 345.8 GHz for CO(3--2).
At the distances of our galaxies, these correspond to spatial scales of several kiloparsecs, such that 80-90\% of the total CO flux typically falls within the beam. We derive (small) aperture corrections by assuming inclined exponential disk profiles for the CO with scale lengths and inclinations derived from optical images \citep{saintonge+17}. This approach has been are validated against resolved CO maps in previous studies (\citet{2011A&A...534A.102L, 2014A&A...564A..65B, 2014MNRAS.445.2599B, 2021ApJS..257...43L}).

The aperture corrected fluxes are used to derive total CO luminosities. Ancillary quantities, including SFRs, stellar masses, and galaxy sizes are derived from the xCOLD GASS catalog \citep{saintonge+17}. SFRs are derived by combining ultraviolet (UV) and infrared (IR) tracers to account for both unobscured and dust-obscured star formation \citep{janowiecki+17}.

\subsection{Data Analysis}

To analyze the relationship between SFR surface density ($\Sigma_{\rm SFR}$) and molecular gas surface densities traced by CO lines ($\Sigma_{\text{CO}}$), we used a statistical approach that combines binning with the Markov Chain Monte Carlo (MCMC) method \citep{foreman2013emcee, hogg2010data}. The objective of this analysis is to determine the power-law slope ($\alpha$) in the relation between $\Sigma_{\text{SFR}}$ and $\Sigma_{\text{CO}}$.

For each galaxy, we compute the galaxy-scale SFR and molecular gas surface densities as
\begin{equation}
    \Sigma_{\text{SFR}} = \frac{1}{2}\frac{\text{SFR}}{\pi r_e^2}
\end{equation}
\begin{equation}
    \Sigma_{\text{CO}} = \frac{1}{2}\frac{L^\prime_{\text{CO}}}{\pi r_e^2}
\end{equation}
where $r_e$ is the optical half light radius, and the factor of one half removes SFR or CO emission arising outside this radius. The use of the optical radius for both calculations is justified given the close correspondence between CO and optical light profiles \citep[e.g.][]{leroy+09}

We characterize the $\Sigma_{\rm SFR}$ and $\Sigma_{\rm CO}$ relation using the functional form:
\begin{equation}\label{eq1}
\Sigma_{\text{SFR}} = C \, \Sigma_{\text{CO}}^{\alpha},
\end{equation}
where $\alpha$ is the slope (or index), and $C$ is a normalization factor. We compared the $\Sigma_{\text{SFR}}$ with three different CO line surface densities within the same galaxy sample: CO(1--0), CO(2--1), and CO(3--2). This approach contrasts with previous studies, which often use measurements from different galaxy samples due to observational limitations at varying redshifts. By analyzing a single, uniform sample, a direct comparison of K--S slopes derived from different CO transitions is ensured, minimizing systematic uncertainties. This allows us to isolate how the power-law slope of the molecular K--S relation varies depending on the excitation conditions of the selected molecular gas tracer. 

\subsection{Binning Method}

To analyse the relationship between molecular gas and star formation surface densities, we used the binning method by dividing the data into five intervals based on ($\Sigma_{\text{CO}}$) for each CO transition, following an approach similar to that used in previous studies (e.g., \citealt{leroy2013molecular}). The bins were constructed to be evenly spaced in $\log(\Sigma_{\text{CO}})$, ensuring a uniform sampling of the CO surface density dynamic range. As a result, the number of galaxies per bin varies depending on the intrinsic distribution of $\Sigma_{\text{CO}}$ in the sample. Within each bin, we calculate the mean and standard deviation of both $\Sigma_{\text{SFR}}$ and $\Sigma_{\text{CO}}$, along with the associated errors. This technique reduces observational scatter and highlights underlying trends. The binned data were plotted alongside the raw data to better visualize trends.

\subsection{Fitting the K--S Relation Using MCMC}

To fit the slopes of the $\Sigma_{\rm SFR}-\Sigma_{\rm CO}$ relations, we convert Equation \ref{eq1} to a logarithmic form: 
\begin{equation}\label{eq2}
\log \Sigma_{\text{SFR}} = \alpha \log \Sigma_{\text{CO}} + \log C + \epsilon(s)\,
\end{equation}
where $\alpha$ and $C$ are as in Equation \ref{eq2}, and the additional $\epsilon$ term allows for an intrinsic scatter around a perfect power law relation. We assume this intrinsic scatter follows a normal distribution, parameterized by width $s$, and we fit for $s$ along with $\alpha$ and $C$.

Bayesian parameter estimation was performed using the Markov Chain Monte Carlo (MCMC) method implemented in the \textsc{emcee} package \citep{foreman2013emcee}. The fitting procedure consists of the following steps: (i) \textit{Log-transform the data}: since the K–S relation follows a power law, we take the logarithm of both $\Sigma_{\text{SFR}}$ and $\Sigma_{\text{CO}}$; (ii) \textit{Define the likelihood function}: assuming normally distributed uncertainties, we use a Gaussian log-likelihood function; (iii) \textit{Initialize the MCMC walkers}: we make initial guesses for $\alpha$, $\log C$, and $s$ and allow the sampler to explore the parameter space; and (iv) \textit{Extract best-fit parameters}: we determine the final values for $\alpha$, $s$, and $\log C$ from the median and 16th/84th percentiles of the posterior distributions.

The MCMC procedure produces posterior samples from which we derive best-fit parameters and their confidence intervals. Specifically, we report the median value of the K–S slope, with uncertainties defined by the 16th and 84th percentiles of the posterior distribution.   The likelihood function for the MCMC sampling accounts for errors in the $x$ and $y$ axes, ensuring that uncertainties in both CO line and SFR surface densities are appropriately propagated into the final fit.

In this work, we do not apply a CO-to-H$_{2}$ conversion factor $\alpha_{\text{CO}}$, as our analysis is based on the observed CO surface brightness ($\Sigma_{\text{CO}}$) rather than on derived molecular gas masses. Our aim is to compare how different CO transitions trace star formation activity under consistent observational conditions, rather than to quantify the total molecular gas content. By using $\Sigma_{\text{CO}}$ directly, we avoid the need to choose a specific $\alpha_{CO}$ value for the fitting. 

When these luminosity–SFR relations are physically interpreted as gas mass–SFR relations, this approach is equivalent to assuming a constant $\alpha_{\text{CO}}$. In this sense, while variations in the conversion factor could in principle affect the absolute normalization of the Kennicutt--Schmidt relation, they would influence all $\text{CO}$ transitions in the same direction and not alter the relative ordering or slope comparisons between them. Moreover, most literature K--S studies adopt a fixed $\alpha_{\text{CO}}$ or fit the $\text{CO}$ luminosities directly, making our choice of a constant conversion factor consistent with previous work and ensuring a direct comparison in Figure~\ref{fig4} and Table~\ref{Table3}.

\begin{figure*}
    \centering    \includegraphics[width=.95\textwidth]{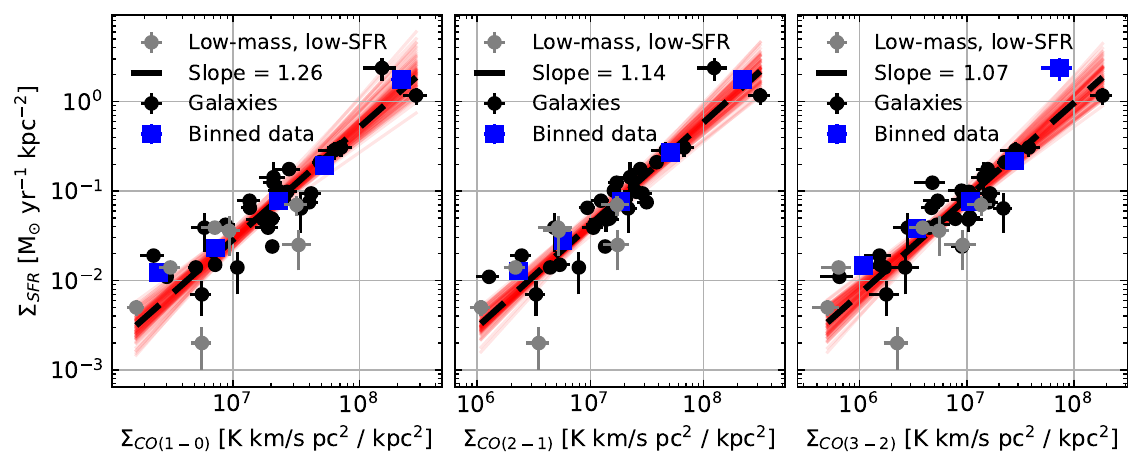}
    \caption{Relationship between SFR surface density ($\Sigma_{\text{SFR}}$) and molecular gas surface density ($\Sigma_{\text{CO}}$) for the full galaxy sample, as traced by CO(1--0) (left panel),  CO(2--1) (middle panel), and  CO(3--2) (right panel). The Black dashed line indicates the best-fitting power law relation, and the thin red lines show 100 randomly selected fits from the MCMC chain. Individual galaxies (black circles) are used in the fit. The gray circles represents the low-mass and low-SFR galaxies that are excluded to form our subsample shown in Figure \ref{fig3}}. Binned data (blue squares) are shown for illustrative purposes. \label{fig2}
\end{figure*}

\begin{figure*}
    \centering    \includegraphics[width=.95\textwidth]{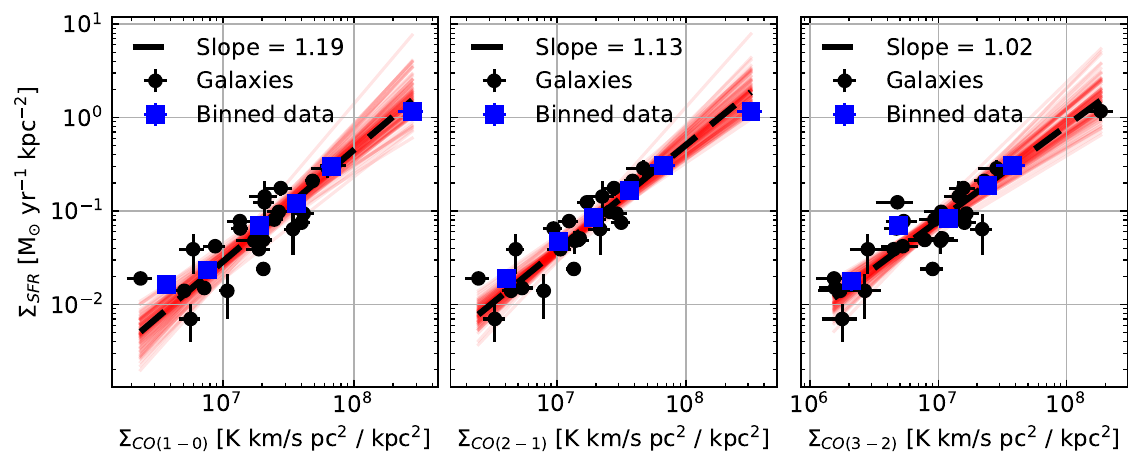}
    \caption{SFR surface density as a function of molecular gas surface density for a subsample excluding low-mass and low-SFR galaxies. Data are shown for CO(1--0) (left panel), CO(2--1) (middle panel), and CO(3--2) (right panel). Individual galaxies (black circles) are fit with the black dashed line. Blue squares (binned data) are shown for illustrative purposes. \label{fig3}}
\end{figure*}

\section{Results} \label{sec:results}

In this section, we present the results of our analysis on the relationship between $\Sigma_{\text{SFR}}$ and $\Sigma_{\text{CO}}$ for the different CO transitions. 
Figures~\ref{fig2} and~\ref{fig3} show the $\Sigma_{\rm SFR}-\Sigma_{\rm CO}$ relations obtained for CO(1--0), CO(2--1), and CO(3--2). 

\subsection{$\Sigma_{\rm SFR}-\Sigma_{\rm CO}$ relations across different CO transition lines}

We conducted our analysis on two samples: the full sample of 36 galaxies with stellar mass ranging from \( 4 \times 10^9 \, M_\odot \) to \( 2 \times 10^{11} \, M_\odot \) and SFR ranging from  \( 8 \times 10^{-2} \, \mathrm{M_\odot\,yr^{-1}} \) to \( 25 \, \mathrm{M_\odot\,yr^{-1}} \) along with a subsample excluding low-mass (Stellar Mass $<10^{10}$ M$_{\odot}$) - low-SFR (SFR $<1$ M$_{\odot}$\,yr$^{-1}$) galaxies.  Figure~\ref{fig2} presents the relationship between $\Sigma_{\rm SFR}$ and the molecular gas surface density ($\Sigma_{\rm CO}$) for the full sample as traced by CO(1--0), CO(2--1) and CO(3--2). A clear positive correlation is observed between $\Sigma_{\text{SFR}}$ and $\Sigma_{\text{CO}}$, and this repeats across all three lines. The data is well represented by a power law relationship, as shown by the linear trend in the plots. To visualize the results we:
(i) plot the observed data points with error bars; (ii) overlay binned data points for clarity; (iii) display 100 randomly selected fits from the MCMC chain to illustrate fit uncertainties; then (iv) plot the best-fit power-law as a dashed black line, with the corresponding K--S slope $\alpha$ indicated in the legend. The binned data points lie very close to the best-fit line across all CO transitions, further supporting the strength and consistency of the fitted relation.

The power-law slopes for the full sample are $\alpha = 1.26 \pm 0.12$ for CO(1--0), $\alpha = 1.14 \pm 0.09$ for CO(2--1),  $\alpha = 1.07 \pm 0.10$ for CO(3--2).

These slopes are shallower than the typically observed slope of $\sim 1.4$ in the K--S relation relating $\Sigma_{\text{SFR}}$ to the total gas surface density, suggesting that the relationship between $\Sigma_{\text{SFR}}$ and $\Sigma_{\text{CO}}$ is closer to linear. While the slope for CO(1--0), $\alpha = 1.26 \pm 0.12$ is not strictly equal to one, it remains significantly flatter than the canonical total gas relation, and the slopes for CO(2--1) and CO(3--2) are even closer to one. This difference implies that the relationship between star formation and molecular gas may differ fundamentally from that of atomic or total gas. This aligns with the works referenced in the introduction, which find similar results.

To investigate potential sampling biases, we repeated our analysis on a restricted, more homogeneous sample, which excludes low-mass and low-SFR galaxies (this subsample is indicated by black boxes in Figure~\ref{fig1}).
The results for this subsample analysis are shown in  Figure~\ref{fig3}. Similar to the full sample, the results present a strong positive correlation between $\Sigma_{\text{SFR}}$ and $\Sigma_{\text{CO}}$ for all three CO transitions. However, the scatter around the best-fitting lines appears to be reduced compared to our full sample. 

The subsample was selected specifically to investigate whether stellar mass acts as a confounding variable in our derived fits. This check was necessary because, due to a selection effect within our main dataset, objects with SFRs less than $1$ M$_\odot$ yr$^{-1}$ are exclusively found among galaxies with low stellar masses. As shown in Figure~\ref{fig1}, this distribution presented a potential bias where stellar mass could inadvertently influence the observed trends at lower SFRs. By analyzing a subsample that excludes these low-SFR, low-mass galaxies, we aimed to assess the robustness of our results against such a bias.

The slopes from the subsample are $\alpha = 1.19 \pm 0.14$ for CO(1--0),  $\alpha = 1.13 \pm 0.11$ for CO(2--1) and $\alpha = 1.02 \pm 0.11$ for CO(3--2). These values are even closer to one compared to those obtained for the full sample, suggesting that the relationship between star formation and molecular gas is more linear in higher-mass, higher-SFR galaxies. 

A slope close to unity in the $\Sigma_{SFR} - \Sigma_{mol}$ implies that the star formation efficiency (SFE), defined as the ratio of star formation rate to gas mass, remains roughly constant as a function of molecular gas surface density ($\Sigma_{mol}$. This constant SFE is characteristic of the molecular gas being fully dominated by giant molecular clouds (GMCs) with self-similar internal properties (e.g., density and free-fall time), potentially indicating a more direct and predictable conversion of molecular gas into stars. Such a near-linear scaling is often interpreted as evidence for star formation being governed by processes within these GMCs structures, rather than by global galactic properties.

The similarity in the slopes before and after excluding low-mass, low-SFR galaxies suggests that these systems do not significantly bias our main results. If these galaxies strongly biased the relation, their removal would be expected to steepen the slope, as they tend to populate the lower end of both $\Sigma_{\text{CO}}$ and $\Sigma_{\text{SFR}}$, potentially flattening the observed trend. Instead, the fact that the slope becomes slightly more linear upon their removal reinforces that they contribute more to scatter than to systematic bias. However, the slightly lower slopes in the refined sample indicate that low-mass galaxies may slightly increase the overall $\Sigma_{\text{SFR}}$ - $\Sigma_{\text{CO}}$ slopes (making the relationship steeper), possibly due to differences in their star formation efficiency or molecular gas content. The use of a more homogeneous sample, therefore, minimizes the influence of galaxies with distinct properties, allowing for a clearer representation of the core relationship between  $\Sigma_{\text{SFR}}$ and $\Sigma_{\text{CO}}$.

\subsection{Statistical Validation}

To assess the robustness and significance of the derived molecular K-S law slopes, we applied several statistical tests.

First, we tested the residuals from each best-fit line for normality using the Shapiro-Wilk and Anderson-Darling tests. In all cases, the residuals are consistent with a Gaussian distribution, supporting the assumptions of our linear regression model. The Shapiro--Wilk test returned p-values between $0.18$ and $0.69$, indicating no evidence against normally distributed residuals at the $5\%$ significance level. The Anderson--Darling statistics ($A^2 = 0.45$, $0.26$, and $0.33$ for the CO(1--0), CO(2--1), and CO(3--2) transitions) are all below the $15\%$ critical value, implying $p > 0.15$ in each case.

Second, we performed 50,000 bootstrap resamplings of our data making Gaussian perturbations to the SFRs, disk sizes, and CO luminosities based on observational uncertainties, and re-derive the molecular KS-slopes for all lines in each realization. The distribution of the resulting slopes confirms that the MCMC-derived slope uncertainties are robust.

\begin{figure}
    \centering
    \includegraphics[width=0.99\linewidth]{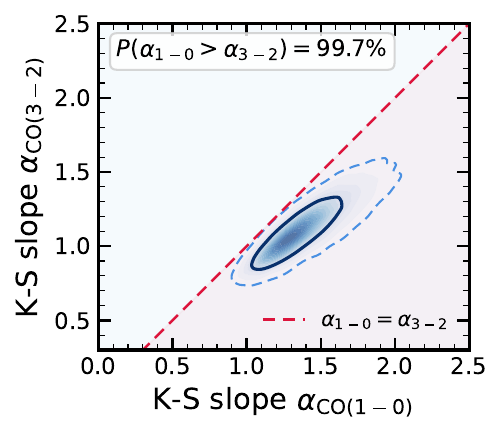}
    \caption{Joint distribution of K--S slopes for the CO(1--0) and CO(3--2) lines across 50,000 bootstrap realizations of our data. In each realization the observed quantities (SFR, CO luminosities, and half-light radius) were perturbed according to their measurement uncertainties, and the K--S relation was re-fitted using orthogonal distance regression (ODR). The contours indicate the $68\%$ (solid) and $95\%$ (dashed) highest-density regions of the bootstrap distribution. The red dashed line marks the equality relation $\alpha_{\mathrm{CO(1-0)}}=\alpha_{\mathrm{CO(3-2)}}$. Realizations below this line correspond to $\alpha_{\mathrm{CO(1-0)}}>\alpha_{\mathrm{CO(3-2)}}$.} 
    \label{fig:bootstrap}
\end{figure}

We note that, because we the $y$-variable ($\Sigma_{\rm SFR}$ is shared across all three $x$-variables ($\Sigma_{\rm CO}$ for each CO line) used in our fits, the uncertainties in our slopes are covariant, and error-bars on the individual $\alpha$s do not capture the significance by which the K-S slope differs for the three CO lines. To illustrate this, Figure~\ref{fig:bootstrap} shows the distribution of bootstrapped slope pairs for CO(1--0) and CO(3--2), highlighting the covariant slope uncertainties. The bootstrap test allows us to quantify the significance of the trend in slope. We find that CO(1--0) has a steeper $\alpha$ than CO(3--2) in $99.7\%$ of realizations, and a steeper $\alpha$ than CO(2--1) in $90.7\%$ of realizations.

Taken together, our validations confirm that the trends in slope with CO transitions are real.  The strongest comparison, between the CO(1--0) and CO(3--2) relations, shows that the CO(1--0) slope is larger in 
$99.7\%$ of bootstrap realizations, corresponding to a significance of approximately $3\sigma$.

\begin{figure*}
    \centering    \includegraphics[width=0.8\textwidth]{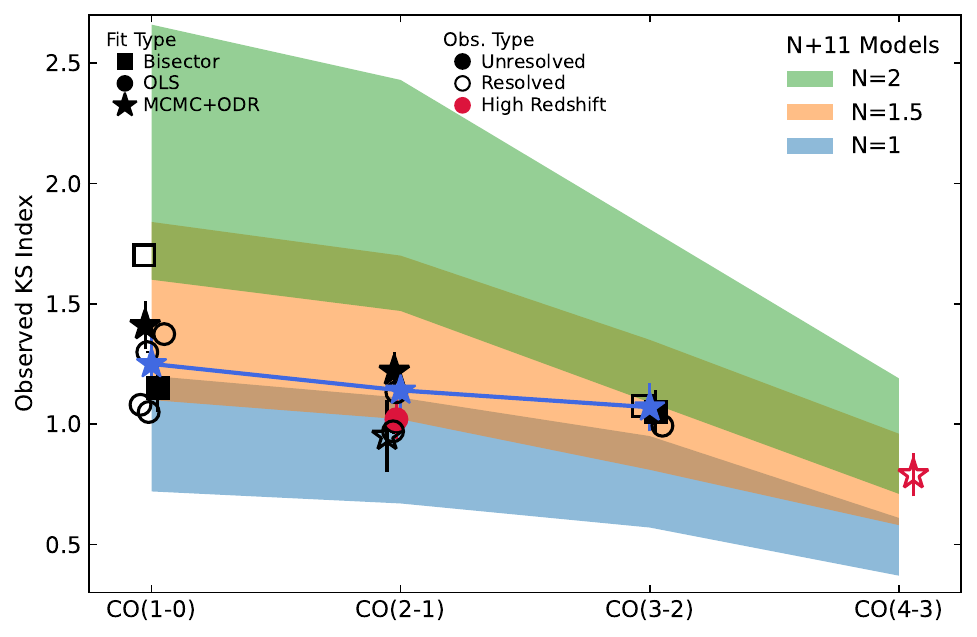}
        \caption{Observed molecular Kennicutt-Schmidt (K--S) index as a function of CO transition from CO(1--0) to CO(4-3). The shaded regions represent theoretical predictions from \citet{narayanan2011kennicutt} for different indices of the underlying (volumetric) SFR density-molecular gas density relation: \( N = 2 \) (green), \( N = 1.5 \) (orange), and \( N = 1 \) (blue). The blue star markers correspond to our data. The remaining data points represent different fitting methods: bisector (squares), OLS (circles), and MCMC + ODR (stars). Filled markers indicate unresolved sources, open markers denote resolved sources, and red markers represent high-redshift sources. Our observed K--S index decreases with increasing CO transition and generally falls in the range of the \( N = 1.5 \) model.
 \label{fig4}}
\end{figure*}

\section{Discussion} \label{sec:discussion}

This study investigates how the excitation conditions of differing molecular gas tracers influence the slope of the molecular K--S relation. We used an uniform sample of nearby galaxies across three CO transition lines, CO(1--0), CO(2--1), and CO(3--2), to examine how each line traces different components of the molecular interstellar medium (ISM) and how these differences influence the observed relation between gas surface density and SFR. Our approach was intentionally comparative, focusing not on a single transition but on how the slope and scatter of the molecular K--S relation change across CO transitions from lower to higher excitation lines. Each transition emphasizes slightly different conditions of the molecular gas, from the colder and more diffuse phase captured by the  CO(1--0) line to the denser and more excited gas highlighted by CO(3--2). The idea was to test whether these lines lead to consistent or diverging star formation relations when applied to the same galaxy sample under the same observational conditions. 

We found that the molecular K--S slopes decrease systematically from CO(1--0) to CO(3--2), with values ranging from $1.26 \pm 0.12$ to $1.07 \pm 0.10$, accompanied by variations in intrinsic scatter. This trend highlights how the choice of molecular gas tracer affects the interpretation of star formation laws.

The higher transitions, which are more closely linked to denser and warmer gas, yield a more linear relation between gas and SFR. These results suggest that the molecular K--S relation is dependent on the physical conditions traced by the chosen molecular gas tracer. The following sections compare our findings to theoretical expectations and previous observational studies and interpret the slope variations in the context of ISM structure and star formation physics. 

\subsection{Comparison with previous studies and theoretical models}

For comparison with our results, we have compiled molecular K--S law indices from a variety of literature sources in Table \ref{Table3} and Figure \ref{fig4}. Table \ref{Table3} includes a compilation of K--S law slopes, along with the CO line analyzed, the number of galaxies, the redshift and fitting method, as well as information on whether the galaxy sample used is resolved or not. Figure \ref{fig4} shows the slopes as a function of CO transition, along with theoretical predictions from the models of \citet{narayanan2011kennicutt}. 

Our data suggest a systematic decline in the K–S slope with increasing CO transition. A bootstrap analysis indicates that the CO(1--0) slope exceeds the CO(3--2) slope in $99.7\%$ of realizations, providing evidence that the molecular K--S relation becomes flatter toward higher--J CO transitions.

A similar pattern is suggested in the literature, where several studies report steeper slopes for CO(1--0) compared to CO(2--1) or CO(3--2), though with some scatter. In particular, studies such as \citet{lamperti+20} and \citet{yajima+21} find differences in slope across CO transitions consistent with our measurements, supporting the idea that higher-J lines trace more directly star-forming gas. By contrast, the results from \citet{morokuma-matsui2017} exhibit a notably larger difference between the CO(1--0) and CO(3--2) slopes compared to most other studies, although the overall trend of a steeper CO(1--0) slope relative to CO(3--2) is still present.
While our study focused on CO(1--0) to CO(3--2), the extension to CO(4–3) in Figure~\ref{fig4} reinforces that the choice of CO transition significantly affects the derived slope of the K--S relation. Higher-J transitions such as CO(3--2) and CO(4–3) tend to exhibit a more linear relation with $\Sigma_{\text{SFR}}$, suggesting that they trace denser, warmer gas more directly involved in star formation. In contrast, lower-J lines like CO(1--0) capture molecular gas in a broader range of conditions, including more diffuse or quiescent components, which can result in steeper slopes. This is also consistent with near-linear correlations found between high-J CO luminosity and SFR \citep{Liu_2015}. This behavior supports the idea that higher excitation CO lines provide a cleaner tracer of star-forming gas, although it should be noted that mechanisms not directly tied to star formation (e.g. AGN) can contribute to the excitation of CO in some environments \citep{salome2023star}.

The models developed by \citet{narayanan2011kennicutt} provide a theoretical framework for interpreting observed star formation relations. These simulations focus on idealized, gas-rich galaxy discs at redshift $\sim 2$ using hydrodynamic simulations combined with molecular line radiative transfer calculations. A key feature of their approach is the incorporation of an underlying \textit{volumetric} Schmidt relation, where the SFR volume density scales with gas volume density as $\rho_{\rm SFR} \propto \rho_{\rm gas}^N$.
The Schmidt index N characterizes how sensitively star formation responds to gas density: when \( N=1\), the relation is linear, meaning the star formation density increases proportionally with gas density. A value of \( N = 1.5 \) implies a superlinear dependence, where increases in gas density lead to more than proportional increases in star formation. In more extreme cases, \( N = 2 \) indicates a very steep, quadratic relation in which star formation becomes dramatically more efficient as gas density rises, typical of compact starburst regions. They explore different values of the index $N$ ranging from 1 to 2, and then use their simulations to predict observable surface density relations, i.e., $\Sigma_{\rm SFR} \propto \Sigma_{\rm gas}^{\alpha}$, by modeling the CO excitation under varying physical conditions. One of the main results of their work is that even with a fixed volumetric slope (e.g., \( N = 1.5 \)), the observed surface density slope $\alpha$ can vary depending on which CO transition is used. This is due to differential excitation effects: higher-J CO transitions (e.g., CO(3--2)) preferentially trace warmer, denser gas more directly associated with ongoing star formation, and thus yield shallower observed slopes, while lower-J transitions (e.g., CO(1--0)) also trace more diffuse or quiescent molecular gas, leading to steeper slopes. The \citet{narayanan2011kennicutt} models, therefore, provide a physically motivated way to connect the observed K--S relation to the underlying physics of star formation.

Our measurements, together with the literature points shown in Figure~4, are broadly consistent with the region predicted by \citet{narayanan2011kennicutt} for an underlying Schmidt index of $N = 1.5$. In contrast, the data are less consistent with the predictions for other values of $N$. Within this framework, the apparent decrease in the $\Sigma_{\rm SFR}$--$\Sigma_{\rm CO}$ slope from CO(1--0) to CO(3--2) is also naturally reproduced by the models. This suggests that the variation in slope across CO transitions reflects differences in excitation conditions rather than intrinsic changes in the star formation efficiency.

These findings demonstrate that multi-line CO observations, when interpreted alongside theoretical models, can yield meaningful constraints on the physical mechanisms governing star formation in galaxies.

A comparison with the results summarized in Table~\ref{Table3} shows general consistency across studies. Nonetheless, variation in slope values (e.g., the slope of $0.79 \pm 0.09$ reported by \citet{nagy2023}) indicates the potential influence of factors such as redshift and sample selection. The high-redshift data from \citet{nagy2023} included in Figure~\ref{fig4} exhibit a lower K–S index; however, this result is based on a higher-$J$ CO transition, where theoretical work suggests intrinsically flatter slopes. By contrast, the high-redshift measurements from \citet{freundlich2019phibss2} are broadly consistent with the low-redshift CO(2--1) results presented here, indicating no compelling evidence for evolution in the K--S slope with redshift when similar CO tracers are used.

\begin{table*}[ht]
    \centering
    \caption{Comparison of molecular K--S law slopes and fitting methods across different CO transitions reported in this work (matched galaxy sample in CO(1--0), CO(2--1), and CO(3--2)) and previous studies using varying CO lines and galaxy samples. The columns are (1) the reference for the reported molecular K--S slope, (2) whether the fit was performed using resolved regions of galaxies or galaxy-integrated values, (3) the (approximate) redshift of the sample used, (4) the CO line used, (5) the number of individual galaxies in the sample (for resolved studies the number of data points being fit is higher), (6) the slope of the molecular K--S relation and its uncertainty (when available), and (7) the fitting method used to derive the K--S slope.}  
    \label{Table3} 
    \begin{tabular}{ccccccc}
        \hline
        Reference & Resolved? & Redshift & CO line & N & K--S law slope & Fitting Method$^{\rm a}$ \\
        \hline
        This work & no & 0.02-0.05 & CO(1--0) & 36 & 1.26 $\pm$ 0.12 & MCMC/ODR \\
        This work & no & 0.02-0.05 & CO(2--1) & 36 & 1.14 $\pm$ 0.09 & MCMC/ODR \\
        This work & no & 0.02-0.05 & CO(3--2) & 36 & 1.07 $\pm$ 0.10 & MCMC/ODR \\
        \citealt{keenan2025arizona} & no & 0.02-0.05 & CO(1--0) & 120 & 1.41 $\pm$ 0.10 & MCMC \\
        \citealt{keenan2025arizona} & no & 0.02-0.05 & CO(2--1) & 120 & 1.22 $\pm$ 0.08 & MCMC \\
        \citealt{nagy2023}$^{\rm b}$ & yes & $\approx$ 1 & CO(4-3) & 2 & 0.79 $\pm$ 0.09 & DLS \\
        \citealt{pessa2021} & yes & 0 & CO(2--1) & 18 & 1.01 $\pm$ 0.01 & OLS (binned data) \\
        \citealt{yajima+21} & yes & 0 & CO(1--0) & 28 & 1.299 $\pm$ 0.005 & OLS \\
        \citealt{yajima+21} & yes & 0 & CO(2--1) & 28 & 1.132 $\pm$ 0.004 & OLS \\
        \citealt{lamperti+20} & no & $\approx$ 0 & CO(1--0) & 87 & 1.15 $\pm$ 0.10 & Bisector \\
        \citealt{lamperti+20} & no & $\approx$ 0 & CO(3--2) & 87 & 1.05 $\pm$ 0.09 & Bisector \\
        \citealt{freundlich2019phibss2} & no & 0.5-0.8 & CO(2--1) & 60 & 1.02 $\pm$ 0.08 & OLS \\
        \citealt{lin2019} & yes & 0 & CO(1--0) & 14 & 1.05 $\pm$ 0.01 & ODR \\
        \citealt{utomo2017edge} & yes & $\approx$ 0 & CO(1--0) & 52 & 1.08 & OLS (binned data) \\
        \citealt{leroy2013molecular} & yes & 0 & CO(2--1) & 30 & 0.95 $\pm$ 0.16 & MCMC (Blanc+2009) \\
        \citealt{morokuma-matsui2017} & yes & $\approx$ 0 & CO(1--0) & 3 & 1.374 & OLS \\
        \citealt{morokuma-matsui2017} & yes & $\approx$ 0 & CO(1--0) & 3 & 1.701 & Bisector \\
        \citealt{morokuma-matsui2017} & yes & $\approx$ 0 & CO(3--2) & 3 & 0.993 & OLS \\
        \citealt{morokuma-matsui2017} & yes & $\approx$ 0 & CO(3--2) & 3 & 1.076 & Bisector \\
        \hline
    \end{tabular}
    \begin{flushleft}
    $^{\rm a}$ Acronyms/abbreviations used: MCMC - Markov Chain Monte Carlo; ODR - Orthogonal Distance Regression; DLS - Levenberg-Marquardt algorithm, damped least squares regression; OLS - ordinary least squares. \\
    $^{\rm b}$ We use the slope measured at 800 pc scale.
    \end{flushleft}
\end{table*}

\citet{keenan2025arizona} also derived molecular K--S relations based on the AMISS dataset, using a larger sample of galaxies but only CO(1--0) and CO(2--1). Their reported slopes of 1.41±0.10 for CO(1--0) and 1.22±0.08 for CO(2--1) are somewhat steeper than our corresponding values, but are consistent with our finding of a decrease in slope for higher energy CO lines. As both studies use similar methodologies and CO line catalogs, the differences in slope likely arise from differences in sample selection. Repeating the analysis of \citet{keenan2024arizona} using only the galaxy sample considered here produces slopes consistent with our results; inclusion of a larger number of high-$\Sigma_{\rm SFR}$ galaxies in the \citet{keenan2024arizona} analysis resulted in their steeper slopes. This reinforces the idea that sample selection affects the slope derived to some extent. Notably, this steepening effect is consistent with predictions from the \citet{narayanan2011kennicutt} model, which suggests that the inclusion of galaxies with high star formation surface densities leads to systematically steeper K--S relations.

\subsection{Implications of variations in K--S Slopes}

The decreasing trend in the K--S slope from CO(1--0) to CO(3--2) suggests that higher CO transitions, which trace denser and warmer gas, arise in conditions more directly tied to star formation activity. 
The CO(1--0) line, which traces more diffuse and colder gas, may be less directly coupled to the immediate star formation process, leading to a steeper slope.

The variation observed in previous studies, as compiled in Table~\ref{Table3}, demonstrates that sample selection is a clear factor influencing the reported K--S slopes. Indeed, comparisons within our own work (i.e. between the subsample and the full sample) further underscore how sample characteristics can notably affect the derived relation. Beyond sample selection, however, Figure~\ref{fig4} and Table~\ref{Table3} indicate that other individual factors, such as spatial resolution or the choice of fitting method, do not obviously explain the full range of scatter among the reported K--S measurements. While literature often discusses distinct trends between resolved and unresolved scales (e.g. \citep{Bigiel_2008, Leroy_2008, leroy2013molecular, schruba2011molecular}), Figure~\ref{fig4} reveals a considerable degree of overlap in the observed K--S index values from these different observational approaches across the various CO transitions. Similarly, the choice of fitting method, as illustrated in Figure~\ref{fig4}, does not significantly impact the overall trend, suggesting it is not a primary factor. What this compilation instead emphasizes is that the broad scatter in reported K--S slopes likely arises from a complex interplay of numerous factors. A key strength of our study, therefore, is its homogeneous analysis across all three CO lines within a consistent galaxy sample. By controlling for many of these variables, our work robustly constrains the intrinsic variation in the K--S slope as a function of CO line, which aligns closely with theoretical models such as \citet{narayanan2011kennicutt}.

This trend further supports the interpretation that CO(3--2), with its higher critical density and excitation temperature, more effectively traces the dense molecular gas directly involved in star formation. The near-linear slope we observe for CO(3--2) mirrors findings from studies using other dense gas tracers, such as HCN \citep[e.g.,][]{Gao_2004, Liu_2015, jimenez-donaire+19, neumann+23}, where linear correlations between line luminosity and SFR are usually reported. Although HCN traces gas at even higher densities than CO(3--2), our results reinforce the broader view that star formation is primarily governed by the amount of dense gas, and that higher-J CO lines serve as better proxies for this star-forming component of the ISM compared to lower-J transitions.

\begin{table*}
\centering
\caption{Slope, intrinsic scatter, and intercept f the molecular Kennicutt–Schmidt relation for the full and subsample galaxy sets. The relation fitted is $\log(\Sigma_{\rm SFR}) = \alpha\, \log(\Sigma_{\rm CO}) + \log C$, where $\Sigma_{\rm SFR}$ and $\Sigma_{\rm CO}$ have units of ${\rm M}_\odot\,{\rm yr}^{-1}\,{\rm kpc}^{-2}$ and $({\rm K\,km\,s}^{-1}{\rm \,pc}^{2})\,{\rm kpc}^{-2}$ respectively. Symmetric uncertainties are the mean of upper and lower bounds.}
\label{Table2}
\begin{tabular}{lcccc}
\textbf{CO Transition} & \textbf{Sample} & \textbf{Slope $\alpha \pm \delta\alpha$} & \textbf{Intrinsic Scatter $\sigma \pm \delta\sigma$ (dex)} & \textbf{Intercept $\log C \pm \delta \log C$} \\
\hline
\multirow{2}{*}{CO(1--0)} 
& Full      & $1.26 \pm 0.12$ & $0.26 \pm 0.05$ & $-10.32 \pm 0.87$ \\
& Subsample & $1.19 \pm 0.14$ & $0.22 \pm 0.05$ & $-9.89 \pm 1.04$ \\
\hline
\multirow{2}{*}{CO(2--1)} 
& Full      & $1.14 \pm 0.09$ & $0.21 \pm 0.04$ & $-9.39 \pm 0.67$ \\
& Subsample & $1.13 \pm 0.11$ & $0.16 \pm 0.04$ & $-9.34 \pm 0.76$ \\
\hline
\multirow{2}{*}{CO(3--2)} 
& Full      & $1.07 \pm 0.10$ & $0.24 \pm 0.05$ & $-8.57 \pm 0.69$ \\
& Subsample & $1.02 \pm 0.11$ & $0.19 \pm 0.05$ & $-8.21 \pm 0.77$ \\
\hline
\end{tabular}
\end{table*}

To better quantify the variations in both slope and scatter, Table \ref{Table2} presents the fitted values of the molecular K--S slope ($\alpha$), the intrinsic scatter ($s$) and the intercept ($\log{C}$) for both the full sample and the high-mass subsample. 

Interestingly, while the slope becomes more linear for higher transitions, the intrinsic scatter does not always follow the same trend. We observe the lowest scatter for CO(2--1) ($\sigma = 0.16 \pm 0.04$ dex in the subsample), suggesting this line may provide the tightest correlation with star formation activity in that regime. In contrast, CO(3--2) shows a slightly higher scatter ($\sigma = 0.24 \pm 0.05$ in the full sample and $0.19 \pm 0.05$ in the subsample), potentially reflecting greater sensitivity to ISM conditions or observational limitations at higher excitation transitions. 

The intrinsic scatter is slightly lower for our high-mass, high-SFR sub-sample across all transitions, suggesting these galaxies form a more homogeneous population. This supports the idea that limiting the range in galaxy properties can help isolate clearer relationships in the star formation law. Although this subset excludes lower-mass systems, our results remain consistent with the established trend that, across the full galaxy population, stellar mass and SFR are positively correlated. Additionally, the intercepts increase toward higher-J transitions, which is expected since higher-J CO lines trace denser, warmer gas and may appear more closely connected to star-forming regions. Focusing on a more homogeneous subset of galaxies reduces the influence of variations in star formation efficiencies and molecular gas content, thereby revealing more consistent trends across the sample.

\section{Conclusions} \label{sec:conclusion}

In this study, we investigated the molecular Kennicutt–Schmidt relation using a uniform sample of 36 nearby galaxies from the AMISS survey, focusing on the CO(1--0), CO(2--1), and CO(3--2) transitions. Our primary goal was to assess how the choice of molecular gas tracer influences the observed correlation between gas surface density and star formation rate surface density.

We find evidence for a small but systematic decrease in the derived molecular K--S slope with increasing CO transition: $\alpha = 1.26 \pm 0.12$ for CO(1--0), $1.14 \pm 0.09$ for CO(2--1), and $1.07 \pm 0.10$ for CO(3--2). This trend is further supported by our literature survey of molecular K--S slopes measured using a variety of CO lines, and is reproduced in numerical simulations.
The increasingly linear correlations when using tracers of higher critical density and excitation temperature are consistent with star formation being more directly linked to the densest component of the molecular ISM \citep{Gao_2004}. Conversely, the super-linear slope for CO(1--0) suggests that a significant fraction of molecular gas lies in a relatively quiescent state, especially in galaxies exhibiting low gas and SFR surface densities.

Our results (illustrated in Table~\ref{Table2}) indicate that CO(2--1) may strike an optimal balance between tracing dense gas and maintaining observational stability, yielding the tightest and most reliable relation. This supports its use as a preferred diagnostic for identifying star-forming gas in nearby galaxies.

Although a similar trend can be inferred from previously published K--S slopes, significant scatter is observed across studies, likely due to differences in sample selection, resolution, and fitting methodologies. By analyzing all three CO transitions within the same galaxy sample, our study eliminates many of the possible biases introduced by variations in galaxy properties, allowing us to isolate the intrinsic effect of gas excitation on the K--S slope.

Comparisons with previous work demonstrate that sample selection and size can influence derived slopes. However, even with a consistent galaxy set, we observe slope variations across CO transitions, highlighting the dominant role of tracer choice over sample composition.

Comparing our results to theoretical models from \citet{narayanan2011kennicutt}, we find that the scaling between \textit{volumetric} SFR and the molecular gas densities has a power-law index of $N\sim1.5$, implying that star formation becomes increasingly efficient in denser regions. However, the decreasing K--S slope with increasing $J$ does not necessarily imply a change in the underlying star formation efficiency per unit gas mass. Instead, it reflects how different CO transitions trace varying components of the molecular gas: higher-$J$ lines are sensitive to denser and more excited gas, and their luminosity responds non-linearly to total gas mass due to variations in excitation conditions. This interpretation is consistent with the framework of \citet{narayanan2011kennicutt}, where differential excitation and thermalization effects lead to flatter observed SFR–$L_\mathrm{CO}$ relations at high $J$, even if the intrinsic SFR–$\rho_\mathrm{gas}$ relation remains unchanged.

While sub-galactic (resolved) studies have reported different slopes compared to global (unresolved) analyses, often attributing this to spatial scale effects, our results suggest that much of the variation arises from a combination of sample selection, resolution, redshift, and fitting method. The absence of a single dominant factor underscores the value of our homogeneous, controlled approach, which reveals systematic and physically meaningful trends across CO transitions.

We also find that low-mass galaxies subtly steepen the $\Sigma_{\text{SFR}}$–$\Sigma_{\text{CO}}$ relation. Using a refined, more homogeneous subsample focused on galaxies with consistent stellar content and star formation activity allowed us to minimize outliers and reduce intrinsic scatter, yielding clearer trends across transitions.

Overall, our findings emphasize the importance of tracer selection, sample uniformity, and methodological consistency in interpreting the molecular K--S relation. Future studies should extend this approach to high-redshift systems using consistent galaxy samples, although this presents challenges due to the increased difficulty of low-$J$ CO observations at those epochs. Our work reinforces the view that star formation is fundamentally linked to the dense, excited phases of the molecular ISM, and that higher-$J$ CO lines offer more direct insights into this star-forming gas.

\subsection*{Acknowledgments}
This research was conducted and partially funded as part of the Max-Planck Institute for Astronomy Summer Internship Program. VdGGS would like to acknowledge the collaborative and supportive environment created by colleagues at MPIA, which significantly contributed to the completion of this project. Additional data analysis and paper preparation were carried out at the Rhodes Centre for Radio Astronomy Techniques and Technologies (RATT), where VdGGS is a PhD student. VdGGS acknowledges the use of the Rhodes University computing clusters for data processing and analysis. 
 The MeerKAT telescope is operated by the South African Radio Astronomy Observatory, which is a facility of the National Research Foundation (NRF), an agency of the Department of Science and Innovation (DSI). VdGGS’s research is supported by the South African Research Chairs Initiative of the DSI/NRF (grant No. 81737). Sincere thanks go to Prof. Oleg Smirnov and Dr. Ian Heywood for their guidance and support, particularly for encouraging participation in the internship during the early stages of the PhD journey. VdGGS also thanks Dr. Kabelo Kesebonye and Dr. Eric K. Maina for their encouragement during a period of uncertainty about pursuing this publication.
 RPK thanks Prof. Daniel Marrone and Dr. Garrett Keating for their guidance in the creation and execution of AMISS, and Dr. Fabian Walter for his input in the development of this project. This paper makes use of data collected by the UArizona ARO Submillimeter Telescope and the UArizona ARO 12-meter Telescope, the IRAM 30m telescope, and the Sloan Digital Sky Survey. The UArizona ARO 12-meter Telescope on Kitt Peak and the UArizona ARO Submillimeter Telescope on Mt. Graham are operated by the Arizona Radio Observatory (ARO), Steward Observatory, University of Arizona. IRAM is supported by INSU/CNRS (France), MPG (Germany) and IGN (Spain). Funding for the SDSS and SDSS-II was provided by the Alfred P. Sloan Foundation, the Participating Institutions, the National Science Foundation, the U.S. Department of Energy, the National Aeronautics and Space Administration, the Japanese Monbukagakusho, the Max Planck Society, and the Higher Education Funding Council for England. The SDSS website is \url{www.sdss.org}.

\bibliographystyle{aasjournal}
\bibliography{bib}
\end{document}